\documentclass[12pt]{article}

\newcommand{\beq}{\begin{equation}}
\newcommand{\eeq}{\end{equation}}
\newcommand{\bea}{\begin{eqnarray}}
\newcommand{\eea}{\end{eqnarray}}
\def\nc{noncommutative}

\def\tjk{\theta_{jk}}
\def\dj{\partial_{j}}
\def\om{\omega}
\def\no{\nonumber}

\begin{document}

\title{\bf Noncommutative Quantum Mechanics and rotating frames}
\author{ {H.R. Christiansen} \\
{\normalsize {\it Centro Brasileiro de Pesquisas F\'{\i}sicas}},
{\small CBPF - DCP} \\
{\normalsize {\it Rua Dr. Xavier Sigaud 150, 22290-180,
Rio de Janeiro, Brazil}} \\ \\ {\small and} \\ \\
F.A. Schaposnik \thanks{Associated with CICPBA, Argentina}
\\
{\normalsize {\it Departamento de F\'\i sica, Universidad Nacional
de La Plata}}\\
{\normalsize {\it C.C. 67, 1900 La Plata\ Argentina}}
}

\date{\tt To appear in Phys. Rev. D}

\maketitle
\begin{abstract}
We study the effect of noncommutativity of space on the physics
of a quantum interferometer located
in a rotating disk in a gauge field background. To this end, we develop a
path-integral approach which allows defining an effective action
from which relevant physical quantities can be computed as in the usual
commutative case. For the specific case of a constant magnetic field,
we are able to compute, exactly, the noncommutative Lagrangian and the
associated shift on the interference pattern for any value of $\theta$.
\end{abstract}

\section{Introduction and Results} The interest in noncommutative space,
recently aroused in connection with developements in string theory
\cite{CDS}-\cite{SW}, rapidly spread on other domains going  from
Quantum field theories  and Quantum mechanics to Condensed matter
physics \cite{MRS}-\cite{GJPP} (See \cite{DN} for a complete list
of references).  Concerning quantum mechanical problems, since
noncommutative physics can be connected with the  dynamics of
charged particles in a magnetic field (the Landau problem), many
interesting results have been presented, going from the
Aharonov-Bohm effect to the Quantum Hall effect
\cite{BS}-\cite{GJPP}.

The purpose of the present work is two fold. On the one hand,
we want to discuss the specific  quantum mechanical problem of a
charged particle in a rotating disk, in the presence of an electromagnetic
field, when space is the anticommutative plane. This is an interesting
problem related to the Aharonov-Bohm effect, relevant to the physics of
superconducting interferometers.

On the other hand, we want to develop
a simple procedure to handle, within the path-integral approach,
noncommutative quantum mechanical problems. The idea is to provide the
Feynman path-integral alternative to the wave equation approach
developed in \cite{BS}-\cite{M}. This last approach is based in taking
into account noncommutativity of the base space by using the so called
$*$ product when the potential in the Hamiltonian acts on the wave function.
Now, at the Hamiltonian level, this amounts to an appropriate shift in the
coordinate dependence of the potential
(and no change in momenta) so that, finally,
noncommutativity is encoded in the shifted potential through the
noncommutative parameter  $\theta_{ij}$. Our approach
starts precisely at this point and makes use of the Feynman recipe
for constructing, in phase space, the transition amplitude $Z$ for a quantum
system in noncommutative space as an integral
over trajectories. Now, for simple (quadratic both in  $p$ and $x$)
potentials, one can integrate over momenta ending with $Z$ written
as a path-integral over $x$, with an effective action
where noncommutativity manifests just through the
parameter $\theta_{ij}$ appearing
in the effective action.

\vskip .5cm

Let us summarize the main results of our investigation. Concerning
the path-integral treatment of a general noncommutative planar
system described by a quantum Hamiltonian $\hat H$, we construct
the transition amplitude between given initial and final states in
the form
\beq Z = \int {\cal D}\vec p {\cal D}\vec x \exp
(\int_{t_i}^{t_f}\!\!\!dt   \left(i(\vec p \dot{\vec x} -
H^\theta_{eff}(\vec p,\vec x))\right) \label{intZZ} \eeq
with the effective Hamiltonian $H^\theta_{eff}$ given by
\beq
 H^\theta_{eff} = \frac{1}{2m}{\vec p}^{\,2}+ V(\vec x -
\vec{\tilde p}) \label{introH}
 \eeq
and
\beq
\tilde p_i = \frac{1}{2} \theta \varepsilon_{ij} p_j
\eeq
with $\theta$  the parameter characterizing noncommutativity. The
only change with respect to the usual Feynman formula in
ordinary space is that the (classical) potential $V$ appears
with its argument shifted due to the presence of $\theta$.

Depending on the precise form of the potential, one should be
able to integrate over
$\vec p$ in (\ref{intZZ}) ending with the Lagrangian version
of the transition amplitude,
\beq
 Z = \int {\cal D}\vec x \exp\left(i\int_{t_i}^{t_f}\! \!\! dt L_{eff}\right)
 \label{inteff}
 \eeq
where $L_{eff}$ can be computed in close form or after some
approximation depending of the type of potential.

Formul{\ae} (\ref{intZZ})-(\ref{inteff}) can be easily applied to
the planar system we are interested in discussing, namely that of
charged particles in a disk rotating with angular velocity $\omega$,
subject to a constant magnetic field $2B$. In this case $L_{eff}$ can
be computed closely, taking the form
\beq
L_{eff} =\frac{1}{2m}
\frac{\left( m v_i - (q B\left(1 +  qB\theta /2\right)
+ m \omega ) \varepsilon_{ij} x_j \right)^2}
{m\omega\theta + (1+ qB\theta /2)^2}
- \frac{1}{2m}(q B x_i)^2 .
\label{introfull}
\eeq
This is an exact result to all orders in $\theta$ which reduces to\
the classical Lagrangian in the $\theta \to 0$ limit which
corresponds
to ordinary space. Also, it reproduces to first order in $\theta$
the
approximate result presented in \cite{jab2}.

Finally, using $L_{eff}$ we shall be able to analyse the
interference pattern of charged particles when a two-slit device
is put on a rotating disk in a gauge field background. We thus obtain a
close expression to all orders in $\theta$ for the phase shift of the 
particle wave-functions. Particularly interesting is the result
that we obtain taking the accelerated interferometer as a rotating
SQUID (superconducting quantum interference device). As an example, for a
magnetic field confined to a thin center hole in the SQUID, the
phase shift between two charged particles takes to first order in
$\theta$ the form
\beq
\Delta \Phi^{0,1} =
 \left(- \frac{2\pi\Delta d}{\lambda} +
\frac{\lambda\Delta d^3}{4\pi} m^2\omega^2 +2 m\om S_w\right)
(1- m\om\theta).
\label{b=0intro}
\eeq
where $d$ is the distance from the source to the detector,
$\Delta(d)$ the difference between the two paths, $S_B$ the
area where the magnetic flux is different from zero and
$S_\omega$ the corresponding one for an effective flux related
to the rotation effect. We discuss the difficulties concerning the
experimental settings in view of the present bounds on $\theta$.

The plan of the paper is the following. After discussing the
classical system in Section 2, we present the path-integral
treatment of the quantum problem in Section 3 and discuss the
quantum interference device in section 4. Finally we present
in Section 5
a summary of our results.
\section{The classical system}
Let us start by defining
 the Moyal $*$-product of functions on the noncommutative plane,
\begin{equation}
\left.(f*g)(x)  = \exp\left(\frac{i}{2} \theta_{ij} \partial_{x_i}
\partial_{y_j}
\right)
f(x)g(y)\right\vert_{y=x}
\label{1}
\end{equation}
Here $\theta_{ij} = \theta \varepsilon_{ij}$ ($i,j=1,2$), with $\theta$
a real parameter
with dimensions of $({length})^2$. The Moyal bracket is then defined as
\begin{equation}
\{f(x),g(x)\} = (f*g)(x) - (g*f)(x)
\label{2}
\end{equation}
Now, for  $f=x^1$ and $g=x^2$, eq.(\ref{2}) takes the form
\begin{equation}
\{x^1,x^2\} = i\theta
\label{3}
\end{equation}
which can be connected,
using the Moyal-Weyl correspondence, with the operator algebra
approach to noncommutative
quantum mechanics where one starts from the commutation relation
\begin{equation}
[\hat x^1,\hat x^2] = i\theta
\label{4}
\end{equation}

Let us consider a particle with mass $m$ and charge $q$
located in a disc rotating with constant angular velocity
$\omega$, in the presence of a gauge field background  $A_i$. It
is an interesting system since, as we shall see, rotational
effects are connected to magnetic ones, and the rotating disk
introduces topological features equivalent to that resulting from
a confined magnetic flux. In a region of a rotating frame that is
not simply connected, the inertial forces can be cancelled without
completely cancelling the inertial vector potential, and its
presence can be detected in a quantum interference experiment as
with the Aharonov-Bohm effect \cite{carmi}. We shall construct
here the Hamiltonian of such a system and then analyze the
associated quantum problem in noncommutative space. Dynamics of
such a classical system is governed by the Lagrangian
\beq
L= -m \sqrt{g_{ij} \dot x^j\dot x^j} -qA_i\dot{x^i} - V
\label{5}
\eeq
where $V$ is some additional potential.
For nonrelativistic velocities
in an inertial frame, we can write (we take for the moment $c=1$)
\beq
L= \frac 12 m v^2 + q \vec v \cdot \vec A - V
\label{6}
\eeq
We want to discuss the case of
 a constant magnetic field. In the ordinary (commutative) case,
this can be very simply achieved by considering a gauge field of the form
 \beq
 A_i =  \varepsilon_{ijk} B_j x_k
 \label{b1}
 \eeq
and  identifying $2B$ with the constant magnetic field (say in
the $z$ direction) computed from $F_{12}$. Interestingly enough,
one can see that already at the classical level coordinate
noncommutativity can be established, this showing its link with
the presence of a magnetic field. Indeed, consider the large magnetic field
limit (equivalent to small $m$) in which the kinematical momentum
$\vec p = m\vec v$
vanishes so that $\vec p = 0$ should be imposed as a constraint.
One has then to introduce Dirac brackets ending with the result
(see \cite{GJPP},\cite{revJ} for a detailed discussion)
\beq
\{ x^i,x^j\}_{Dirac} = \frac{1}{2qB} \varepsilon^{ij}
\label{Dirac}
\eeq
We see  in this very simple
classical system how noncommutativity arises because of the
presence of the constant magnetic field background, with $\theta$
and $B$ related according to $\theta =
1/(2qB)$.

Let us now consider the same classical problem but in
noncommutative plane. In this case, the appropriate field strength,
 which in fact changes
 covariantly under noncommutative $U(1)$ gauge transformations, should
 be defined as
 \beq
 F_{\mu\nu} = \partial_\mu A_\nu - \partial_\nu A_\mu -iq\{A_\mu,A_\nu\}
 \label{F}
 \eeq
 so that the gauge potential (\ref{b1})  yields  a field strength
 with the $F_{12}$ component in the form
\beq
F_{12} = 2B \left(1+{q\theta} B/2 \right)
\label{f12}
\eeq
As expected, $F_{12}$ coincides, to zeroth order in $\theta$,  with $2B$,
the value of the magnetic field associated with the gauge field (\ref{b1})
in the commutative plane.
The term linear in $\theta$ modifies the commutative result giving, however,
a field strength that is still constant.

It is worthwhile to mention here that,
in general, local quantities in noncommutative electrodynamics are not
gauge invariant and only integrated expressions can be given an
invariant measurable meaning. Of course,  a constant noncommutative
field strength,
is still gauge invariant since for constant $F_{12}$ one has
 $g^{-1}* F_{12}* g = F_{12}$. Furthermore, for general
 field strengths which are just gauge covariant, a
 possibility to work with gauge invariant objects is
provided by Seiberg-Witten mapping \cite{SW}. Indeed, this mapping
connects a noncommutative gauge theory with an ordinary one
formulated in terms of ordinary (not star) products of gauge
fields and an action having an explicit dependence on
$\theta^{ij}$ which acts as a constant background field. Once the
gauge theory is expressed in terms of ordinary gauge fields and of
the background $\theta^{ij}$,  it becomes a theory which is gauge
invariant in the conventional sense with an action from which
gauge invariant electric and magnetic fields can be defined. This
is the strategy adopted in \cite{GJPP}, \cite{revJ} which also
applies to define covariant coordinate transformations \cite{JP}.
In general, the  mapping can be determined by solving the
Seiberg-Witten differential equation order by order in $\theta$. To order
$\theta$, the transformation relating the noncommutative field
strength $F_{ij}$ and the corresponding one, which 
 in the commutative
equivalent theory will be denoted as $F^C_{ij}$,  is
\beq
F_{ij} = F^C_{ij} + \theta_{kl}\left( F^C_{ki}F^C_{lj} -
A^C_k\partial_l F^C_{ij} \right) + \ldots
\label{small}
\eeq
Gauge invariant electric and magnetic fields can then be computed
from $F^C$. 

Since in the present quantum mechanical context one is in general interested
in small $\theta$ effects, Eq.(\ref{small}) is enough to establish a
relationship between the noncommutative field strength and its
commutative counterpart. Moreover, for an Abelian constant $F_{ij}$,
the differential equation can be solved explicitly. The solution
(with boundary condition $F_{ij}(\theta = 0) = F^C_{ij}$) written
in order to have $F^C_{ij}$ in terms of $F_{ij}$ is \cite{SW}
\beq
F^C = F \frac{1}{1 - q\theta F} ,
\label{FF}
\eeq
where for notation simplicity we have supressed indices. Formula
(\ref{FF}) then gives an explicit way to connect magnetic and
electric fields in the noncommutative and its equivalent
commutative theories. As signaled above, gauge-covariant rules for
the transformation of gauge fields under tranformations of
noncommutative coordinates can be defined using the
Seiberg-Witten mapping.

We now come back to Lagrangian (\ref{6}) and write it
in the rotating frame. Using (\ref{b1}) we write
\beq
L= \frac 12 m {\vec v}^2 +  q \vec v \cdot \vec B \times \vec r - V
\eeq
(we ignore the additional potential $V$ in what follows).
In order to write the Lagrangian in the rotating frame (with constant
angular velocity $\omega$ which we take parallel to
$B$) one has just to change $\vec v \to \vec v + \vec\omega\times\vec r$
getting,
\beq
L=\frac 12 m {\vec v}^2 + m \vec v \cdot
(\vec \omega \times\vec  r) +\frac 12m (\vec \omega\times \vec r)^2 +
q \vec r \cdot (\vec v
\times \vec B) +  q \vec B \cdot (\vec r\times
(\vec \omega\times \vec r)) .
\label{laga}
\eeq
Now, if we  define a vector field $\vec {\cal V}$
such that
\beq
\vec {\cal V} = \vec \omega \times \vec r ,
\label{a}
\eeq
so that the canonical momentum reads
\beq
 \vec P= \frac{\partial L}{\partial \vec v} = m\vec v + m\vec {\cal V}
  + q \vec A .
 \label{P}
 \eeq
With this, the Hamiltonian in the rotating frame takes the form
\beq
H= \frac 1{2m} (\vec P- m \vec {\cal V} - q \vec B \times \vec r )^2 -
\frac{m}{2} \vec{\cal V}^2 - {q} \vec {\cal V} \cdot (\vec B \times \vec r).
\label{Ha}
\eeq

\section{The quantum system in noncommutative space:
the path-integral approach}
In order to discuss the quantum mechanical problem, let us start by
noting  that in the noncommutative plane,
the Heisenberg algebra takes the form
(we put $\hbar = 1$)
\begin{eqnarray}
{[\hat x^1, \hat x^2]} &=& i \theta ,\nonumber\\
{[\hat p^1, \hat p^2]} &=& 0, \nonumber
\\
{[\hat x^i,\hat p^j]} &=& i\delta^{ij} . \label{nofun}
\end{eqnarray}
Now, because of (\ref{nofun}), it is not possible to
construct  eigenstates $|x^1,x^2\rangle$ common to $x^1$ and $x^2$ and hence
the definition of a probability density for a
given state $|\psi\rangle$  becomes problematic. However, as noted in
 \cite{BS}-\cite{jab2},
one can find a new coordinate system
\beq
x^i = \hat x^i + \tilde p^i , \ \ \ \ \  p^i = \hat p^i , 
\label{sombrerito}
\eeq
with
\beq
 {\tilde p}^i =  \frac{1}{2} \theta \varepsilon^{ij} \hat p_j ,
\label{tilde}
\eeq
which satisfies the canonical commutation relations
\begin{eqnarray}
{[ x^1,   x^2]} &=& 0,\nonumber\\
{[p^1,  p^2]} &=& 0,\nonumber\\
{[x^i, p^j]} &=& i \delta^{ij} .
\label{normal}
\end{eqnarray}
It is then possible in this new system to define the probability
density associated with a state $|\psi\rangle$
as $|\langle y^1 y^2 | \psi\rangle |^2$. One may then use
 realization (\ref{normal}) to solve specific quantum
mechanical problems.

An alternative approach to investigate noncommutative quantum
mechanical systems is related to the way in which noncommutative quantum
field theories haved been investigated. In this approach, one
starts from the  Schr\"odinger equation with ordinary products replaced
by Moyal $*$-products as defined in (\ref{1}) while
coordinates are treated as in ordinary
space \cite{BS},\cite{M}. We shall follow this last approach and
consider, for definiteness, the simple case in which the system corresponds
to a particle of mass $m$,  in a potential $V(\vec x)$.  The
Schr\"odinger equation for such a system should then be written as
\beq
i \frac{\partial\psi(\vec x,t)}{\partial t} =
\frac{1}{2m}\hat{\vec p}^{\,2} \psi(\vec x,t)+ V(\vec x) * \psi(\vec x,t) .
\label{Sca}
\eeq
Now,
one can eliminate the $*$ product in the potential term
by using \cite{BS}-\cite{M}
\beq
V(\vec x) * \psi(\vec x,t) =   V(\vec x - \vec {\tilde p}) \psi(\vec x,t) ,
\label{corri}
\eeq
an identity that can be proven just by Fourier tranforming the
l.h.s.  Once this is done, the  wave equation reads
\beq
i \frac{\partial\psi(\vec x,t)}{\partial t} =
\frac{1}{2m}\hat{\vec p}^{\,2} \psi(\vec x,t)+
V(\vec x - \vec{\tilde p} ) \psi(\vec x,t)
\equiv \hat H_{eff} \psi(\vec x,t) , 
\label{Scae}
\eeq
which is a ``normal'' (ordinary space)
Schr\"odinger equation
for a system with a modified Hamiltonian $\hat H_{eff}$. One can
make contact between this approach and that referred at the beginning
of this section by noting that a redefinition of coordinates
according to eq.(\ref{sombrerito})  turns eq.(\ref{Scae}) into the
Schr\"odinger equation for the original Hamiltonian but
with coordinates obeying the algebra (\ref{nofun})

\vskip .5cm

We are now ready to investigate the path-integral approach
to the quantum problem in noncommutative space.
To this end,
we shall proceed to the construction of the quantum
transition amplitude using the Feynman integral over trajectories. As
it is well known, this approach replaces the analysis of
the
wave equation for a system with quantum Hamiltonian $\hat H$
by the phase space path-integral $Z$ giving the transition amplitude between
some given initial and final states
\beq Z = \int {\cal D}\vec p {\cal D}\vec x \exp
(\int_{t_i}^{t_f}dt \left(i(\vec p \dot{\vec x} - H(\vec p,\vec
x))\right) \label{ZZ} \eeq
where $H(\vec p,\vec x) = \langle \vec p |\hat H|\vec x\rangle $.

Now, in view of eq.(\ref{Scae}),
one can apply the usual Feynman recipe to the system with Hamiltonian
$\hat H_{eff}$, and write the transition
amplitude in the form
\beq
Z = \int {\cal D}\vec x {\cal D}\vec p \exp \left(i(\vec p \dot{\vec x} -
H_{eff}(\vec p,\vec x))\right)
\label{ZZZ}
\eeq
with $H_{eff}(\vec p,\vec x) = \langle \vec p |\hat H_{eff}|\vec x\rangle $.
It is just in the shifted potential term  in $H_{eff}$  where noncommutativity
manifests. Depending on the form of the potential, which
depends now on $\vec p$
because of the shift (\ref{corri}), the integral over momenta could be
done in close form, leading to a Lagrangian version of $Z$ .

In the case of Hamiltonian (\ref{Ha}), the shift (\ref{corri}) amounts to
\begin{eqnarray}
A_i = -\varepsilon_{ij} B x_j  &\to& -\varepsilon_{ij} B
\left( x_j - \tilde p_j
\right) \nonumber\\
{\cal V}_i = -\varepsilon_{ij} \omega x_j &\to &-\varepsilon_{ij} \omega
\left( x_j - \tilde p_j
\right)
\label{cha}
\end{eqnarray}
As a result, Hamiltonian $H_{eff}$ can
be written in the form
\beq
H_{eff} = \frac{1}{2m}\left(1 + qB \theta/2 \right)^2
       \left(\vec p -  \frac{ q \vec A}{1 + qB\theta/2} \right)^2
+\frac 12\, \omega p^2\ \theta
     -\vec p\cdot \vec \omega \times \vec r.
\label{ham}
\eeq
In the present case, this expression can be used to define
effective mass and charge resulting from
deformation of space at the \nc\ scale \cite{M},
\begin{eqnarray}
m_{eff} &=& \frac{m}{\left(1 + qB\theta/2\right)^2} \nonumber\\
q_{eff} &=& \frac{q}{{1 + qB\theta/2}}.
\end{eqnarray}

Being $H_{eff}$ quadratic in $\vec p$, one can integrate out the
momenta
in (\ref{ZZZ}), this yielding to the
Lagrangian version of the path-integral $Z$. The answer is
\beq
 Z = \int {\cal D}\vec x \exp\left(i\int \! dt L_{eff}\right)
 \label{eff}
 \eeq
where the effective Lagrangian $L_{eff}$ is given by
\beq
L_{eff} =\frac{1}{2m}
\frac{\left( m v_i - (q B\left(1 +  qB\theta /2\right)
+ m \omega ) \varepsilon_{ij} x_j \right)^2}
{m\omega\theta + (1+ qB\theta /2)^2}
- \frac{1}{2m}(q B x_i)^2 .
\label{full}
\eeq
Note that this is the exact expression for the
Lagrangian, to all orders in $\theta$.
As expected, it reduces to the classical one for
$\theta=0$.  It is important to stress that applying
 the \nc\ transformation defined in eq.(\ref{cha})
to the classical Lagrangian (\ref{laga}) {does not}
yield the effective Lagrangian eq.(\ref{full}).
This is due to the fact that the former is obtained
after path-integrating the momenta (this implying that
factors in  the numerator of the shifted Hamiltonian
appear as denominators
in the Lagrangian) and not just by a simple shift in
the $\vec x$ variables.

Up to first order in $\theta$, the effective Lagrangian
can be written, in terms of vector fields,  as
\bea
L_{\theta^{0,1}} & = & \frac 12 m
(v_i +{\cal V}_i+ \frac {2q}m A_i)(v_i+{\cal V}_i) -
\frac 12qm{\tjk} (\dj A_i)
 (v_i +{\cal V}_i+ \frac qm A_i)(v_k+{\cal V}_k) \nonumber\\
&&
- \frac 12m^2 {\tjk} (\dj {\cal V}_i) \left((v_i +
{\cal V}_i)(v_k+{\cal V}_k+
2\frac qm A_k)
+\frac{q^2}{m^2} A_iA_k\right).
\label{1order}
\eea
Written in this way it is instructive to show the structure
of the aproximate \nc\ Lagrangian for a  {generic} case.

Using a three-dimensional notation,
one can further rewrite the first order expression compactly, as
\bea
L_{\theta^1} = &&
-\frac{qm}{4\hbar^2}  \vec\theta \cdot \left((\vec v+\vec {\cal V})
\times \nabla A_i\right)  \left(v_i+{\cal V}_i+\frac qm A_i\right)
\nonumber \\
&&
- \frac{m^2}{4\hbar^2}  \vec\theta \cdot \left(
(\vec v+\vec {\cal V} +2\frac qm \vec A)
\times \nabla {\cal V}_i\right) (v_i+{\cal V}_i) \nonumber\\
&&
- \frac{q^2}{4\hbar^2} \vec\theta \cdot \left(
\vec A \times \nabla {\cal V}_i \right)A_i .
\label{vectorial}
\eea
where we have defined $\vec\theta{~}_i= \epsilon_{ijk}\tjk$.
One can easily see that eq.(\ref{vectorial})
coincides with the approximate (first order in $\theta$)
result derived in
\cite{jab2} for the special case of ${\cal V}=0$.
It should be stressed, however, that eq.(\ref{full}) provides an
 {exact} form for the Lagrangian to be considered in the transition
amplitude $Z$ for the quantum noncommutative model.

\section{The quantum interference device}

We are now in conditions to discuss the quantum dynamics
 of  charged particles in a rotating disk, in the presence
of a gauge field background. In this way,
by studying the interference pattern
of the particles when a two slit device is put on a
rotating disk, we shall be able to determine noncommutative
effects in connection both with
the gauge field and with the non-inertial frame.
As we shall see, both effects interfere each other
and provide a $\theta$ shift
which can be acurately calculated.

Let us start by observing that the phase shift $\Delta \Phi$
between two electrons reaching a detector through different
paths can be computed from the formula
\beq
\Delta \Phi = \Delta \int_{t_i}^{t_f} dt\ L_{eff}
\label{Delta}
\eeq
where $\Delta$ indicates subtraction between both integrals
computed in the interval $({t_i},{t_f})$ that the particle takes
from the source to the detector.
For the particular case of a constant $F_{12}$,
the full Lagrangian eq.(\ref{full}) takes the simple form
\beq
L_{eff}= \alpha v_i^2 +\beta x_i^2
+\gamma v_i \epsilon_{ij} x_j
\eeq
with
\bea
&& \alpha = \frac m{2 f_{\theta}} \nonumber\\
&& \beta = \frac 1{2m} \left( \frac{g_{\theta}^2}{f_{\theta}}
    -q^2 B^2 \right) \nonumber\\
&& \gamma = - \frac{g_{\theta}}{f_{\theta}}\nonumber\\
&& f_{\theta} = 1+\theta(m\omega+qB) + \theta^2
\frac{q^2B^2}{4}\nonumber\\
&& g_{\theta} = m\omega+qB +  \theta \frac{q^2B^2}2.
\eea
Now, the result of the integration in eq.(\ref{Delta})
in terms of the de Broglie wavelegth $\lambda = 2\pi/p$
associated with the particle is
\beq
\Delta \Phi = \frac{2\pi}{\lambda f_{\theta}} \Delta(d)
+ \frac{\lambda}{4\pi}( \frac{g_{\theta}^2}{f_{\theta}}
    -q^2 B^2 )
\Delta(d^3) +  \oint \vec\gamma\ d\vec x
\eeq
where $d$ is the distance from the source to the detector (thus,
$\Delta(d)$ represents the difference between the two paths to the
same point in the detector) and
$\gamma_i= \gamma\epsilon_{ij} x_j$.
While the first two terms do not depend on the flux of
the fields, the last one does. Indeed, the $\gamma$ factor in
the third term
depends on the area where the flux of magnetic and ${\cal V}$
  fields are non-zero.
Suppose that one confines the $B$ flux into a solenoid in the center
of the rotating disk. Then the $B$ part of
the curl$\,\vec\gamma$ flux would be
multiplied by the area of the solenoid while the $\omega$ part by the
area defined by the path difference. In this way
we would have an Aharonov-Bohm effect combined with a
rotational effect. If the solenoid is very thin, and the magnetic
field not too strong, then the relevant phase shift will depend on the
angular velocity. Nevertheless, it must be noted that
these effects are not only summed but also
multiplied each other, (see for example eq.(\ref{vectorial})).

In order to clearly distinguish the \nc\ contributions from those
already present in the ordinary case,
let us analyse the complete first order approximation, given by
\bea
\Delta \Phi^{0,1} &=& - \frac{2\pi\Delta d}{\lambda} +
\frac{\lambda\Delta d^3}{4\pi}  (m^2\omega^2+2qBm\omega)
+2(m\om S_w+ qB S_B) +
\no\\
&& \theta \left( \frac{2\pi\Delta d}{\lambda} (m\om+qB) -
\frac{\lambda\Delta d^3}{4\pi}(m\om+qB)(m^2\omega^2+2qBm\omega) \right.
\no\\
&& {~~~~~} \left. -2 m^2\omega^2 S_w -2 qBm\omega (S_w+S_B) -
q^2B^2 S_B \right) ,
\eea
where $S_B$ and $S_w$ are respectively the areas
where the magnetic flux and $\om$
flux are nonzero (it must be noted that since $S_w$ is defined by
the particle's contour, then it is always $S_B < S_w$).
The last term in the first line represents the
usual Aharonov-Bohm contribution,
while the second one is the rotational analog.
In the third line, we find the
corresponding \nc\ shifts, and their interference becomes apparent.
The other terms do not depend on the topology of the device but we can
also see both the ordinary and their \nc\ counterparts.
Our result shows that the device can be used to exhibit
the \nc\ shift in the Aharonov-Bohm effect, and
introduces a new physical effect
due to the interference of the two potential fields $\vec A$ and
 $\vec {\cal V}$.

\noindent At this order, the $\vec {\cal V}=0$ phase shift is simply
\beq
\Delta \Phi^{0,1}|_{\vec{\cal V}=0} =
-\frac{2\pi\Delta d}{\lambda} (1- qB \theta) - qB S_B(2-qB\theta).
\label{a=0}
\eeq
The accelerated interferometer could be also realized as a
rotating SQUID (superconducting quantum interference device).
In this case, irrespective of the external field
distribution, the particle in the SQUID would not see any $B$ field
as a result of the Meissner effect, and thus no magnetic force would
be measured in the rotating frame.  Nevertheless, there is still
a magnetic flux through
the center of the SQUID, together with that related to $\omega$, this
amounting
to the effect just described.

If on the other hand one has not a strong magnetic field or it
is confined to a thin center hole in the SQUID, then the
$\omega$ field
still affects the particle resulting in a phase shift of the same
nature, also depending on $\theta$ as follows
\beq
\Delta \Phi^{0,1} =
 \left(- \frac{2\pi\Delta(d)}{\lambda} +
\frac{\lambda\Delta (d^3)}{4\pi} m^2\omega^2 +2 m\om S_w\right)
(1- m\om\theta).
\label{b=0}
\eeq
In order to have a measurable \nc\ effect, the first order
contribution should be a measurable fraction of the  $\theta$=0
result, say a 1$\%$. In this case, from eq.(\ref{a=0}), one should
have a strong magnetic field satisfying

\beq
qB \simeq  10^{-2}\theta^{-1} .
\eeq
Similarly, in  eq.(\ref{b=0}) one needs the angular velocity as fast as
\beq
m\om \simeq 10^{-2}\theta^{-1} .
\eeq
Using the current bound for the \nc\ parameter,
$\theta \leq (10 TeV)^{-2}$ \cite{jab},\cite{Harvey},
one finds that both requirements are experimentally hard to realize. One
could perhaps think about an experimental setting involving astronomic
velocities and field-strengths, but it seems to us that it is still
 beyond the current possibilities.

\section{Summary}
We have presented a path-integral approach to noncommutative quantum mechanics
in the plane and discussed how the physics of a rotating interferometer is
affected by the fact that spatial coordinates do not commute.
One advantage of
this approach is that all noncommutative effects are encoded in an effective
Lagrangian which can be used to compute transition amplitudes within
the usual framework provided by Feynman integral over trajectories.
The transition amplitude, originally
written as a path-integral over phase space, eq.(\ref{ZZZ}), can be reduced
to a path-integral over coordinates, eq.(\ref{eff}), with an effective
Lagrangian which includes the effects of noncommutativity of space coordinates.
Using this result, we have investigated a rotating interferometer
device to see how noncommutativity will eventually modify the physics of
a quantum particle in a gauge field background together with the effects
of locating the system in a noninertial frame.
From the associated quantum effective Lagrangian, which for the present
gauge we calculated to all orders in $\theta$,
we were able to analytically compute the phase shift for electrons reaching
the detector through different paths in an exact way. Finally, we
discussed the possible implicancies of noncommutativity on the phenomenology
of a rotating SQUID showing that the current bounds on $\theta$ require
an experimental setting which is far beyond the present possibilities.

\bigskip

\subsection*{Acknowledgments}
H.R.C. dedicates this work to his daughter, Michelle. The authors
wish to acknowledge G.D. Barbosa for valuable collaboration
and discusions in an early stage of this work. Thanks are also due to
S.A. Dias and J. Helayel-Neto for useful comments.
F.A.S  is partially supported by CONICET (PIP 4330/96),
ANPCYT (PICT 03-05179) and CICPBA, Argentina.

\end{document}